\documentclass[12pt,a4paper]{article}
\usepackage{graphicx}
\usepackage{times}
\title{\bf Spectropolarimetry of Beta Lyrae:\\ Constraining the Location of the Hot Spot and Jets}
\author{Jamie R. Lomax$^1$ and Jennifer L. Hoffman$^1$\\
\vspace{1cm}\\
\normalsize $^1$ University of Denver, Denver, CO, USA\\ }
\date{\mbox{}}
\begin{document}
\maketitle
\pagestyle{empty}
%
%
\def\bull{\vrule height .9ex width .8ex depth -.1ex}
\makeatletter
\def\ps@plain{\let\@mkboth\gobbletwo
\def\@oddhead{}\def\@oddfoot{\hfil\tiny\bull\quad
``The multi-wavelength view of hot, massive stars''; 39$^{\rm th}$ Li\`ege Int.\ Astroph.\ Coll., 12-16 July 2010 \quad\bull}%
\def\@evenhead{}\let\@evenfoot\@oddfoot}
\makeatother
%
%
\def\beginrefer{\section*{References}%
\begin{quotation}\mbox{}\par}
\def\refer#1\par{{\setlength{\parindent}{-\leftmargin}\indent#1\par}}
\def\endrefer{\end{quotation}}
%
%
{\noindent\small{\bf Abstract:} 
Beta Lyrae is an eclipsing, semi-detached binary system whose state of active mass transfer can reveal details of the nonconservative evolution of binary stars. Roche lobe overflow has caused the system to evolve to a complex state. A thick accretion disk almost completely obscures the secondary, mass-gaining star while the rapid mass transfer likely drives mass loss through the system's bipolar outflows. Polarimetry can provide important information about the physical structure of complex systems; in fact, the discovery of bipolar outflows in beta Lyrae was confirmed through polarimetry. Here we present results from 6 years of new and recalibrated spectropolarimetric data taken with the University of Wisconsin's Half-Wave Spectropolarimeter (HPOL). We discuss their implications for our current understanding of the system's disk-jet geometry. Using both broadband and line polarization analysis techniques, we derive new information about the structure of the disk, the presence and location of a hot spot, and the distribution of hot line-emitting gas.
}
%
%
\section{Introduction}

\indent\indent Beta Lyrae (hereafter $\beta$ Lyr) is a semi-detached, eclipsing binary star system. The primary star is B6-8 II giant, while the secondary star is likely a B0.5 V main-sequence star (Hubeny \& Plavec 1991). Mass loss from the primary via Roche lobe overflow has caused a thick accretion disk to surround and obscure the secondary. A bipolar flow or jet has also been detected in the system through interferometric (Harmanec et al. 1996) and spectropolarimetric (Hoffman, Nordsieck \& Fox 1998) methods. The jets may originate from a hot spot (Harmanec 2002) caused by the impact of the mass stream on the disk. See Harmanec (2002) for a comprehensive overview of our current understanding of the system.

Since the stars in $\beta$ Lyr are hot, the material in the disk is at least partially ionized. Therefore, the observed polarization is primarily caused by electron scattering. Analyzing the polarization behavior allows us to extract geometric information about the system. Thus we can use polarimetry to probe the location of the hot spot and jets. There is no evidence of a hot spot in $\beta$ Lyr's light curve. However, since the hot spot is a local disruption in the disk, which is the primary scattering region for visible light, we expect it may act as a depolarizing region whose effects would be detectable in the polarization light curve. As long as the hot spot does not lie on the line connecting the centers of mass of the two stars, its depolarizing effect should result in a phase difference between the minimum flux and minimum polarization. 

If the jets originate from the hot spot, we expect to see eclipse effects at phases in the polarized \pagebreak[4] light that do not match those in total light. For the case where the hot spot leads the primary star in phase, the eclipse of the jets would occur after the secondary minimum in scattered visible light.

\section{Observations}
\indent\indent We investigated these predictions by analyzing six years of newly calibrated and previously unpublished spectropolarimetric $\beta$ Lyr data taken with the University of Wisconsin's Half-Wave Spectropolarimeter (HPOL). These data consist of 69 observations taken between the years of 1992 and 1998, the first 29 of which were presented in Hoffman et al. (1998). To calculate the phase of each observation, we have used the quadratic ephemeris given by Harmanec \& Scholz (1993). We used eight observations of $\beta$ Lyr \textit{B}, an unpolarized star physically associated with $\beta$ Lyr and located $45^{\prime\prime}$ away, to establish a new interstellar polarization (ISP) estimate. We subtracted the Serkowski law fit of the mean $\beta$ Lyr \textit{B} polarization spectra from each $\beta$ Lyr observation to correct for ISP effects. 

\subsection{\textit{V} Band Polarimetry}

\begin{figure}[h]
\centering
\includegraphics[width=13cm]{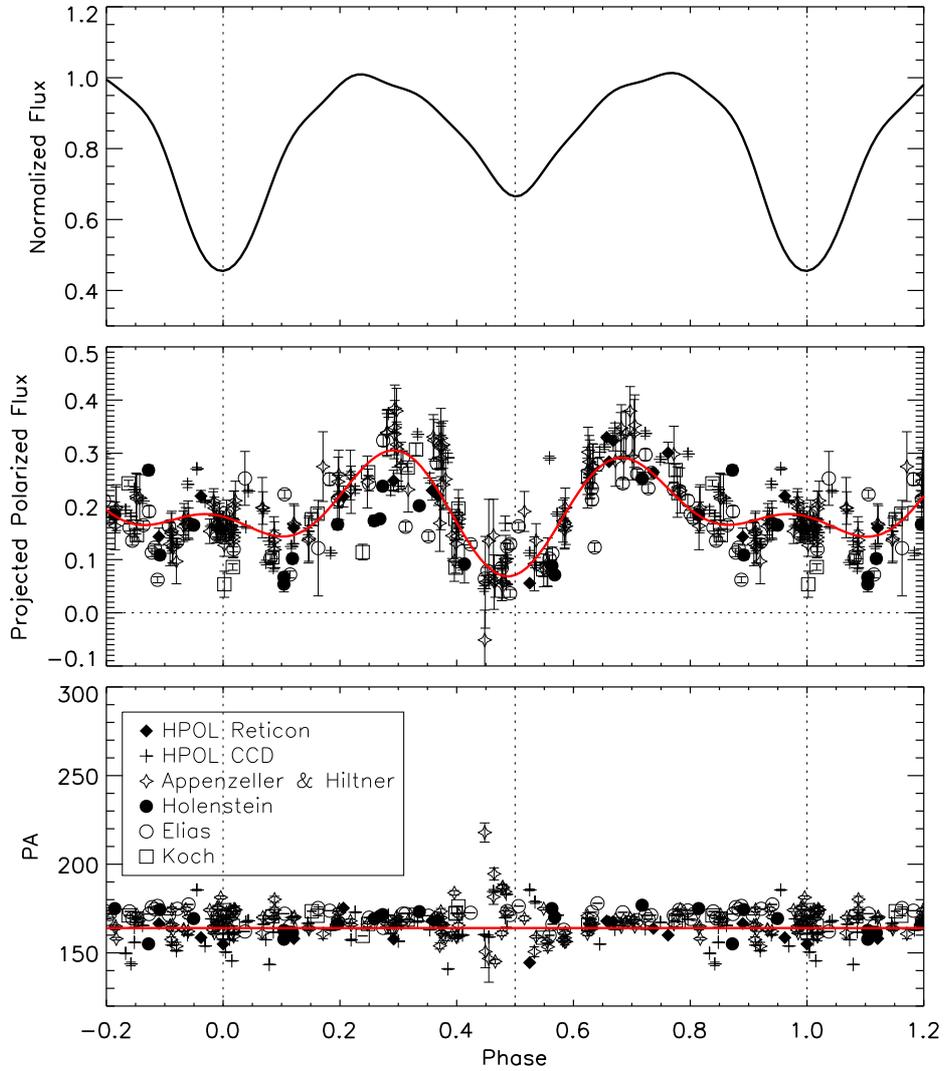}
\caption{The top panel shows the \textit{V} band Fourier fit light curve of $\beta$ Lyr (Harmanec et al. 1996). The middle panel shows the \textit{V} band projected polarized flux curve and a Fourier fit to the data. The bottom panel shows the phase variation of the position angle, with the average position angle of $164^{\circ}$ indicated. Filled diamonds represent HPOL Reticon data, crosses represent HPOL CCD data, open diamonds are data from Appenzeller \& Hiltner (1967), open circles represent data taken with the Flower and Cook Observatory (FCO) by N.M. Elias II, closed circles represent FCO data taken by B.D. Holenstein and open squares represent FCO data taken by R.H. Koch (Elias, Koch \& Holenstein 1996). Both the polarized flux and the position angle show effects just before secondary eclipse that suggest a hot spot influences the disk polarization.\label{fig_1}}
\end{figure}

\indent\indent To look for effects of the hot spot, we constructed a projected polarized flux light curve (Figure \ref{fig_1}). We did this by rotating the Stokes parameters to the system axis ($164^{\circ}$) so that the average \%\textit{u} value is zero and multiplying the rotated \%\textit{q} value by the normalized \textit{V} band light curve (Harmanec et al. 1996). The resulting light curve represents only the scattered light. As shown in Figure \ref{fig_1}, the \textit{V} band data display a shift in secondary minimum when compared to the total light curve. An error-weighted Fourier fit to the projected polarized flux data, represented by a red line in Figure \ref{fig_1}, has a minimum at phase 0.487, while the secondary eclipse occurs at phase 0.5. This matches our prediction of how a hot spot would manifest itself in polarimetric data and provides evidence for its presence. Based on the offset between the secondary eclipse minima, the maximum size of the hot spot across the disk edge visible to the observer is approximately $26.5R_\odot$ at phase 0.487. This is comparable in size to the disk radius, $30R_\odot$, and its large size may be due to contributions from the mass stream to this effect. The large fluctuations in the position angle near secondary eclipse also seen in Figure \ref{fig_1} are consistent with this interpretation, as the hot spot should disrupt the disk edge, causing a randomization of the position angle of light scattered from the hot spot's location. 

We also note that the two maxima in the Fourier-fit \textit{V}-band projected polarized flux curve are slightly unequal in magnitude. This effect decreases at longer wavelengths, demonstrating that the polarized spectrum is bluer before secondary eclipse than after. This effect is consistent with a hot spot model, as we expect the disk temperature to be higher near the spot than on the opposite side.

\subsection{H$\alpha$ Line Polarimetry}

\begin{figure}[h]
\centering
\includegraphics[width=12cm]{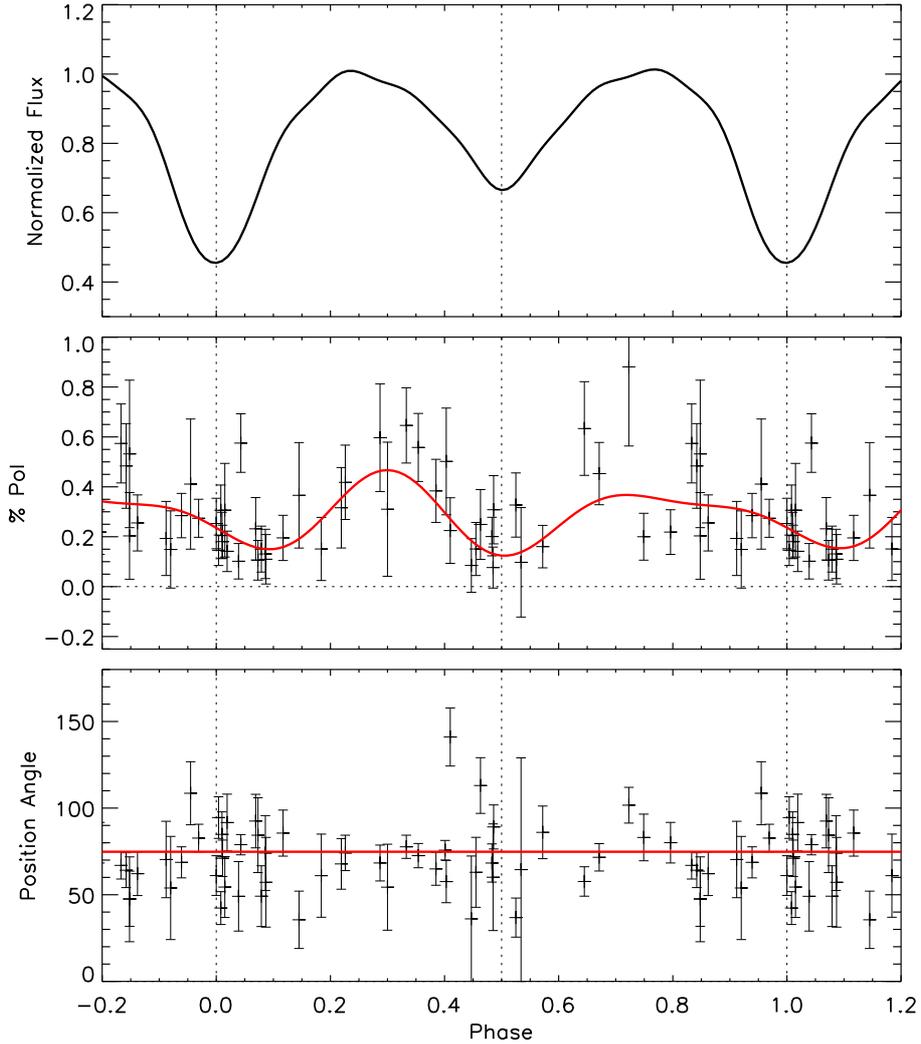}
\caption{The top panel shows the \textit{V} band Fourier fit light curve (Harmanec et al. 1996). The middle panel shows how the polarization of the H$\alpha$ line changes with phase. The curve is a Fourier fit to the data. The bottom panel shows the position angle with the mean value of $75^{\circ}$ indicated. Only HPOL CCD data are shown. The presence of an eclipse centered on phase 0.5 suggests, contrary to previous models of $\beta$ Lyr, that the jets are not centered on the hot spot.\label{fig_2}}
\end{figure}

\indent\indent 
We performed a similar analysis to investigate the polarization behavior of the strong H$\alpha$ (6562.8 \AA) emission line. As Figure \ref{fig_2} shows, its position angle is aligned nearly $90^{\circ}$ away from that of continuum and is therefore likely scattered in the jets. The H$\alpha$ polarization curve shows two eclipses, suggesting that the scattering regions in the bipolar jets are comparable in size to the disk height. The eclipse centered on phase 0.5 indicates, in contradiction to previous models of $\beta$ Lyr, that the jets do not originate from the hot spot. Instead, they are likely centered between the stars near the surface of the secondary, perhaps forming as disk material accretes onto the secondary. However, the large errors and scatter of the data shown in Figure \ref{fig_2} make the exact phase of the secondary eclipse and the jet's location considerably uncertain. 

\pagebreak[4]

\clearpage

\pagebreak[4]
\clearpage

\subsection{New Model of the $\beta$ Lyr System}

\begin{figure}[h]
\begin{minipage}{8.25cm}
\centering
\includegraphics[width=7cm]{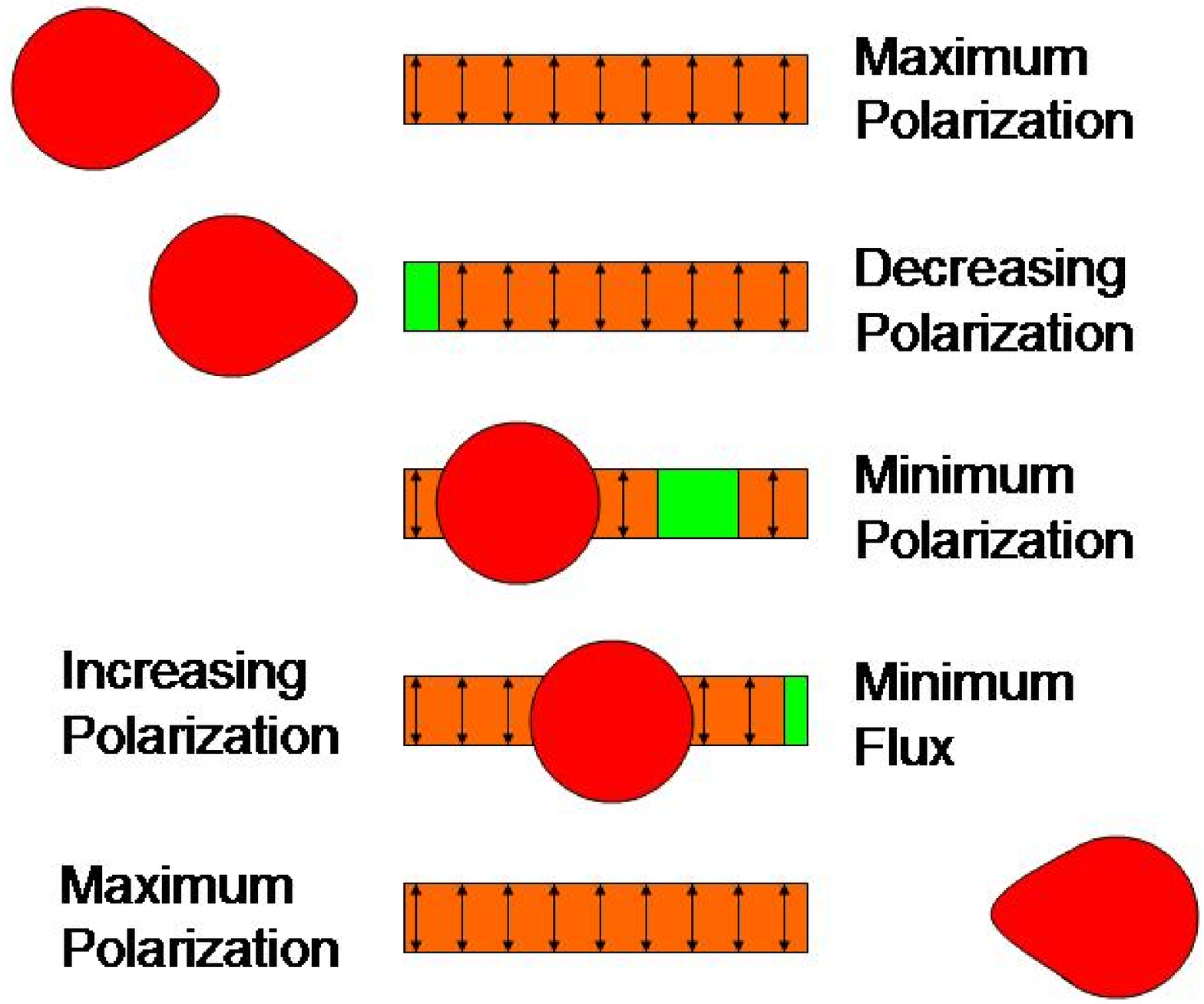}
\caption{This cartoon shows our interpretation of the results presented in Figure \ref{fig_1}. The vectors indicate the relative strength of the polarization signal at the given phases. Minimum polarization occurs before  minimum flux because this is when the disk area eclipsed by the primary and disrupted by the hot spot (highlighted in green) is the largest. This drawing is not to scale. \label{fig_3}}
\end{minipage}
\hfill
\begin{minipage}{8.25cm}
\centering
\includegraphics[width=7cm]{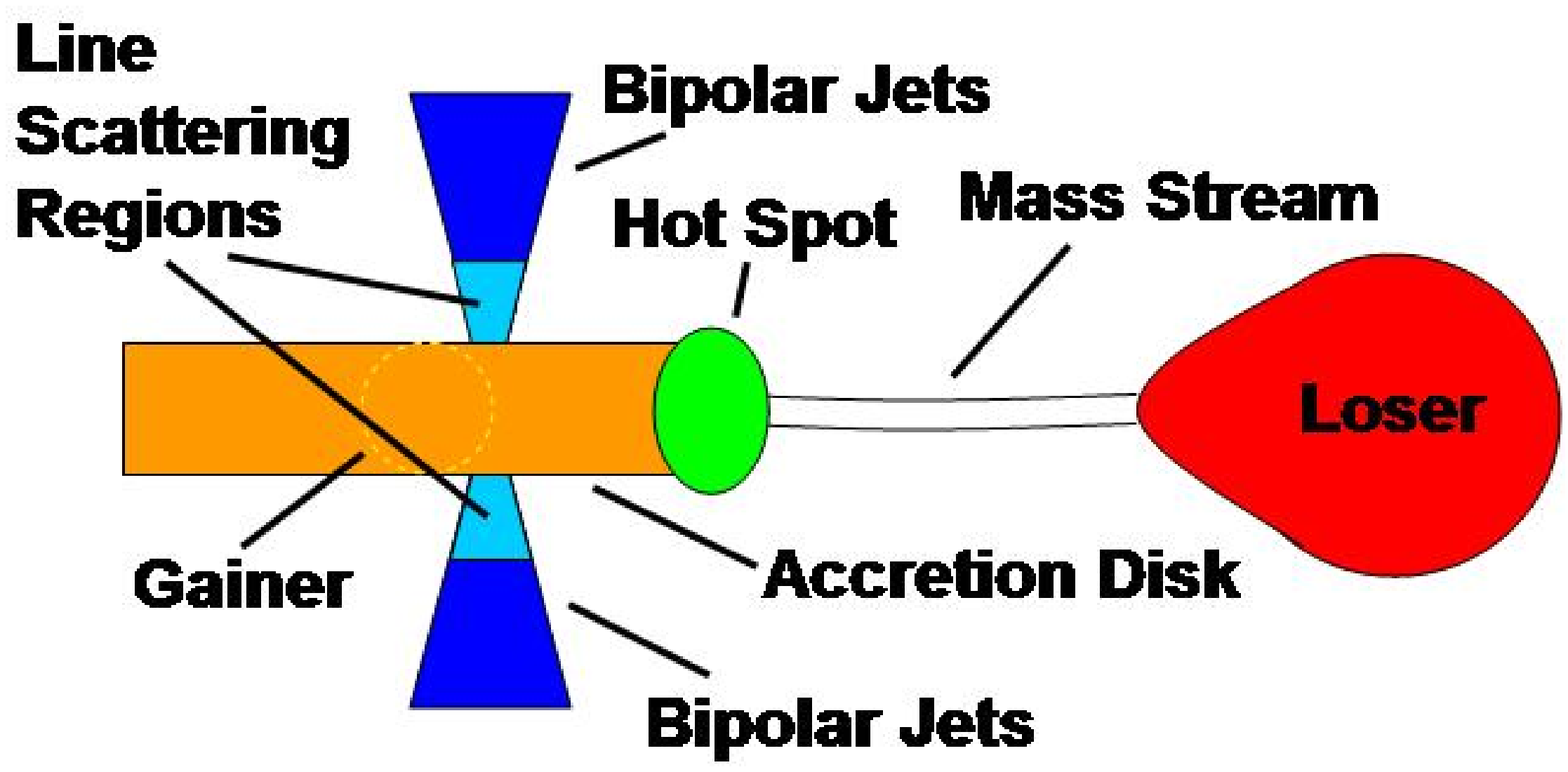}
\caption{This cartoon shows our interpretation of the results presented in Figure \ref{fig_2}. The location of the jets must be near the center of the disk and the sizes of their scattering regions (highlighted in light blue) must be smaller than the height of the primary star to produce the eclipses seen in the H$\alpha$ line polarization. \label{fig_4}}
\end{minipage}
\vspace{-10pt}
\end{figure}

\indent\indent The \textit{V} band and H$\alpha$ polarimetry indicate that a new model of the system is needed. The \textit{V} band result suggests the hot spot begins its transit across the disk before the loser eclipses the disk. As seen in Figure \ref{fig_3}, this model produces a minimum in polarization when the area of the disk covered by the primary star and hot spot is maximized. As minimum flux approaches, the area covered by the hot spot decreases as seen by the observer because it is rotating to the far side of the disk. Figure \ref{fig_4} shows an illustration of our new model that includes the jet location implied by the H$\alpha$ polarimetry results. Improved line polarization observations will help us refine our current model of the system.

\section{Future Work}
\indent\indent Future analysis of these data will provide quantitative constraints on the size and location of the hot spot and line scattering regions. We plan on improving the line polarization results with new 
observations using HPOL at the Mount Lemmon Observing Facility to improve the constraints on the properties of the jets. We will present a more detailed analysis of the results of the line and \textit{U,B,R} and \textit{I} band polarimetry in full in a future paper. In order to detect the mass stream and confirm the existence of the hot spot, we plan to conduct future observations in both the infrared and X-ray regimes. 

While previous studies have detected phase variations on timescales longer than the orbital period of the system, preliminary period analysis of the polarimetric data presented here do not reveal any of these variations. However, this analysis is still ongoing and will be presented in a future paper.

%
%
\section*{Acknowledgements}
\indent\indent Lomax acknowledges the support of the 39th Li\`ege International Astrophysical Colloquium's Organizing Committee in the form of a travel grant, which enabled her to attend this conference. This research is supported by NASA ADP award NNH08CD10C and NSF award AST0807477. A special thanks to Marilyn R. Meade, Petr Harmanec, Kenneth Nordsieck, Richard Ignace and Nicholas M. Elias II for their continued help with this project.
%
%
\footnotesize
\beginrefer
\vspace{-10pt}
\refer Appenzeller, I., \& Hiltner, W.A., 1967, ApJ, 149, 353\\
\refer Elias II, N.M., Koch, R.H., and Holenstein, B.D., 1996, BAAS, 28, 913\\
\refer Harmanec, P., 2002, Astron. Nachr., 323, 87\\
\refer Harmanec, P., Morand, F., Bonneau, D., et al., 1996, A\&A, 312, 879\\
\refer Harmanec, P., \& Scholz, G., 1993,  A\&A, 279, 131\\
\refer Hoffman, J.L., Nordsieck, K.H., and Fox, G.K., 1998, AJ, 115, 1576\\
\refer Hubeny, I., \& Plavec, M.J., 1991, AJ, 102, 1156\\

\endrefer           
\end{document}